\documentstyle[12pt]{article}
\begin{document}
\thispagestyle{empty}

\def\cqkern#1#2#3{\copy255 \kern-#1\wd255 \vrule height #2\ht255 depth
   #3\ht255 \kern#1\wd255}
\def\cqchoice#1#2#3#4{\mathchoice%
   {\setbox255\hbox{$\rm\displaystyle #1$}\cqkern{#2}{#3}{#4}}%
   {\setbox255\hbox{$\rm\textstyle #1$}\cqkern{#2}{#3}{#4}}%
   {\setbox255\hbox{$\rm\scriptstyle #1$}\cqkern{#2}{#3}{#4}}%
   {\setbox255\hbox{$\rm\scriptscriptstyle #1$}\cqkern{#2}{#3}{#4}}}
\def\CC{\mathord{\cqchoice{C}{0.65}{0.95}{-0.1}}}
\def\x{\stackrel{\otimes}{,}}
\def\y{\stackrel{\circ}{\scriptstyle\circ}}
\def\proof{\noindent Proof. \hfill \break}
\def\a{\begin{eqnarray}}
\def\b{\end{eqnarray}}
\def\p{{1\over{2\pi i}}}
\def\Q{{\scriptstyle Q}}
\def\P{{\scriptstyle P}}
\renewcommand{\thefootnote}{\fnsymbol{footnote}}

\newpage
\setcounter{page}{0}
\pagestyle{empty}

\centerline{\LARGE KP Hierarchies, Polynomial and Rational ${\cal W}$}
\centerline{\LARGE Algebras On Riemann Surfaces:}
\centerline{\LARGE A Global Approach.}
\vspace{1truecm} \vskip0.5cm

\centerline{\large F. Toppan}
\vskip.5cm
\centerline{Dipartimento di Fisica}
\centerline{Universit\`{a} di Padova}
\centerline{Via Marzolo 8, I-35131 Padova}
\vskip1.5cm
\centerline{\bf Abstract}
\vskip.5cm
A covariant pseudodifferential calculus on Riemann surfaces, based
on the Krichever-Novikov global picture, is presented. It allows
defining scalar and matrix KP operators, together with their
reductions, in higher genus. Globally defined Miura
maps are considered
and the arising of polynomial or rational ${\cal W}$ algebras on R.S.
associated to each reduction are pointed out. The higher genus NLS
hierarchy is analyzed in detail.
\vfill
\rightline{DFPD 94-TH-58}

\newpage
\pagestyle{plain}
\renewcommand{\thefootnote}{\arabic{footnote}}
\setcounter{footnote}{0}

\section{Introduction.}

\indent

It is very well known that a deep connection exists between the
algebraic geometry and the integrable partial differential equations
of the soliton theory. Such connection has been developed since the
seventees, from the work of, among others, Novikov,
Dubrovin, Matveev, Krichever, \cite{novi}; for a more complete
historical account see \cite{novi2} and references therein. \par
In such a framework, periodic or quasi-periodic solutions are
constructed with the help of the Baker-Akhiezer functions defined on
auxiliary Riemann surfaces. Roughly speaking, the solutions are
obtained from ``spectral" algebraic geometrical data which, in a
generic case, correspond to a bundle over the moduli space
$M_{g,N}$ of smooth algebraic curves of genus $g$ with $N$ punctures.
So for instance, in the case of the KP hierarchy, the solutions are
obtained for the bundles over the moduli space with $1$ puncture
($N=1$).
\par
Another possibility of connecting Riemann surfaces with integrable
equations of non-linear type has been investigated in
\cite{bm}. In this work the KdV equation was covariantized in order
to describe a (euclidean) dynamics over a specified Riemann surface
$\Sigma$, providing in such a way the generalization of the classical
KdV equation on a cylinder (which can be conformally mapped to
describe a dynamics over the punctured complex plane $C^{\star}$).
This work made use of the global description of a
Riemann surface obtained by replacing the
Fourier-Laurent bases with the Krichever-Novikov ones \cite{kn}
relative to the given Riemann surface.\par
In this paper I will reconsider the scheme pioneered in \cite{bm} in
order to formulate in the global Krichever-Novikov framework not only
the KdV or standard Drinfeld-Sokolov type of hierarchies\cite{DS},
but the
whole KP hierarchy, together with its non-standard (i.e.
non-principal in the $gl(n)$-matricial language,\cite{genkdv})
reductions (see also \cite{ara}).
Such reductions lead to the coset hierarchies like the
Non-Linear Schr{\" o}dinger equation, and are associated
to ${\cal W}$ algebras of rational type \cite{{toppan},{toppan2}}.
To reach
this goal one needs to refine the set of ``covariantization rules"
proposed in \cite{bm}:
 basically one needs to
covariantize the whole pseudodifferential operator calculus. \\
A remark is in order: in this framework one has the freedom to choose
among different dynamics:
if we do not force the time variable being identified with the
euclidean time ${\tau}$ on a Riemann surface, then
the evolution equation corresponds to the real time evolution of a
field defined over a punctured Riemann surface
$\Sigma^{\star}$ (here $\Sigma^{\star}=\Sigma\backslash{P_\pm}$, with
$P_\pm$ corresponding to the North and South Pole of the Riemann
surface).  Such dynamics is not uniquely defined: every holomorphic
one-form on $\Sigma^{\star}$
defines its corresponding dynamics. This degeneracy can be removed
with a canonical choice however if the evolution
f{}low
corresponds
to the
f{}low
along the euclidean time, leading to a nice geometrical
interpretation. It should be pointed out however that the
identification
of the evolution parameter with the euclidean time is
mandatory for relativistic theories like the WZNW models
on Riemann surfaces; on the contrary it is not so essential for
non-relativistic systems like those described by the KP hierarchy and
its reductions.
\par
Some useful
results concerning the covariantization of conformal operators (Bol
operators and so on) can be found in \cite{gie}. Recently in a couple
of papers \cite{zuc} the  formulation of Drinfeld-Sokolov hierarchies
on Riemann surfaces and the appearance of ${\cal W}$ algebras in the
light of the Krichever-Novikov framework have been discussed. In these
papers only the principal reductions have been considered.\par The
generalization of the construction here presented to the case of
Riemann surfaces with more punctures can be straightforwardly
performed as well by using the results obtained
in a series of papers by
Schlichenmaier (see e.g; \cite{schli}).
\par
The scheme of the
paper is the following:
In the next section the needed mathematical framework will be
introduced. I will follow the approach and results of \cite{bt},
where in particular some useful results concerning the Heisenberg
representation of the KN algebra were first obtained.
Next, the pseudodifferential calculus will be introduced for Riemann
surfaces. KP hierarchies and their reductions will be conveniently
formulated. The Poisson brackets structure
(leading to ${\cal W}$ algebras)
 will be provided through
(higher genus) Miura maps.
\\ In the fourth section the case of non-principal reductions
of the matrix KP hierarchy on Riemann surfaces will be considered
by analizing in detail as an example the
Non-Linear-Schr{\"o}dinger equation.
The free-field Wakimoto
representation will be realized in the higher genus case and the
appearance of
a rational
${\cal W}$ algebra
(see \cite{toppan}) will be discussed.

{}~\quad\\
\section {Notations and conventions.}

\indent

On a Riemann surface $\Sigma$ of genus $g$
let us consider two
distinguished points $\P_+$ and $\P_-$ and local coordinates $z_+$
and $z_-$ around them, such that $z_{\pm}(P_\pm)=0$. On $\Sigma$ I
will consider bases of meromorphic tensors
which are holomorphic except at $\P_\pm$, in particular of meromorphic
vector fields
$e_I$, functions
$A_I$, 1-differentials $\omega^I$ and quadratic
differentials $\Omega^I$.
Here $I$ is integer or
half-integer according to whether $g$ is even or odd.
The behaviour near $\P_\pm$ is given by
\a
A_I(z_\pm)&=&a{_I^{\pm}}z{_\pm^{\pm I-{g\over 2}}}(1+
{\cal O}(z_\pm))\label{Ai}\\
\omega^I(z_\pm)&=&b{_I^{\pm}}z{_\pm^{\mp I+{g\over 2}-1}}(1+
{\cal O}(z_\pm))(dz_\pm)\label{omegai}\\
e_I(z_\pm)&=&c{_I^{\pm}}z{_\pm^{\pm I-g_0+1}}(1+
{\cal O}(z_\pm)){\partial\over{\partial z_\pm}},\quad \quad
g_0={3\over 2}g\label{ei}\\
\Omega^I(z_\pm)&=&d{_I^{\pm}}z{_\pm^{\mp I+g_0-2}}(1+
{\cal O}(z_\pm))(dz_\pm)^2
\label{Omegai}
\b

For $|I|\leq {g\over 2}$ the definitions (\ref{Ai},\ref{omegai}) must be
modified, because of the Weierstrass theorem. Let us set $A_{g\over
2}=1$, while for $I={g\over
2}-1,\ldots,-{g\over 2}$  the power of $z_-$ is lowered by 1 in
(\ref{Ai}). As  for $\omega^I$ and $I={g\over
2}-1,\ldots,-{g\over 2}$ the power of $z_-$ must be raised by 1 in
(\ref{omegai}) and
$\omega^{g\over 2}$ is set equal to the third kind
differential
\a
\omega^{g\over 2}(z_\pm)= \pm
{1\over {z_\pm}}[1+{\cal O}(z_\pm)](dz_\pm)
\label{omegag/2}
\b
normalized in such a way that the periods around
any cycle are purely
imaginary.

The above bases elements are determined up to numerical
constants due to the Riemann-Roch theorem. So let us set e.g.
$a{^{+}_I}=1$, then the
$a{^{-}_I}$'s are completely determined. One can do the same for
the $c^\pm_I$'s. As for the
other constants they are fixed by the duality relations
\a
{1\over {2\pi i}}\oint_{C_\tau} d\Q~A_I({\scriptstyle Q})
\omega^J({\scriptstyle Q})&=&\delta_I^J\label{duality}\\
{1\over {2\pi i}}\oint_{C_\tau}d\Q~ e_I({\scriptstyle Q})
\Omega^J({\scriptstyle Q})&=&\delta_I^J\nonumber
\b
Here $C_\tau$ denotes a level curve
of the univalent function
\a
\tau({\scriptstyle Q})=Re\int_{Q_0}^Q \omega^{g\over 2}\label{tau}
\b
for a fixed
$\Q_0\in\Sigma$. Of course the integrals of eq.(\ref{duality}) do not
change for a contour which can be continuously deformed to a $C_\tau$.
Henceforth the symbol $\oint$ without any specification will denote
integration
around $C_\tau$.

The Lie brackets of the bases element $e_I$ are
\a
\relax [e_I,e_J]={C^K_{IJ}}~e_K,\quad\quad \label{bracket}
\b
Here and throughout the paper summation over repeated indices
is understood, unless otherwise stated.
The structure constants $C{^K_{IJ}}$ can be calculated from the
constants appearing in the expansion of $e_I$
\a
C^K_{IJ}= \p\oint [e_I,e_J]~ \Omega^K\nonumber
\b
Eq. (\ref{bracket}) defines what
we call the KN algebra over $\Sigma$. The central extension of this
algebra is defined by means of the cocycle
\a
\chi(e_I,e_J)={1\over {24\pi i}}\oint \tilde\chi(e_I,e_J)\nonumber
\b
the integral is over any contour
surrounding $\P_+$, and $\tilde\chi(f,g)$
for any two meromorphic vector
fields $f=f(z){\partial\over{\partial z}}$
and
$g=g(z){\partial\over{\partial z}}$, is given by
\a
\tilde \chi(f,g)=
\bigl({1\over 2}(f'''g-g'''f)-R(f'g-fg')\bigr)dz_+\nonumber
\b
$R$ is a Schwarzian connection.

The relation
\a
\relax [e_I,e_J]={C^K_{IJ}}~e_K+t~\chi(e_I,e_J)
,\quad [e_I,t]=0 \label{bracket'}
\b
defines the extended KN algebra.
\par

In the following the extension
$\hat{\cal A}^\Sigma$ of the commutative
algebra ${\cal A}^\Sigma$ of the $A_I$'s
(generalized Heisenberg algebra),
will be needed; it is
defined by
\a
\relax [A_I,A_J]=\hat k~\gamma_{IJ}, \quad [A_I,\hat k]=0\label{At}
\b
where
\a
\gamma_{IJ}={1\over{2\pi i}} \oint A_I dA_J\label{gamma}
\b
The cocycle $\gamma_{IJ}$ vanishes
for $|I+J|>{g+1}$, as it is easy to verify from
eq.(\ref{gamma}).

In the following one also needs the notations
\a
N_i^J &=& {1\over{2\pi i}}\oint_{a_i} \omega^J \label{Ni}\\
M_i^J &=& {1\over{2\pi i}}\oint_{b_i} \omega^J
\label{Mi}
\b
where $\{a_i, b_i\}, i=1,...,g$ is a basis of homology cycles.
{}From eq.(\ref{gamma}) one gets
\a
dA_I = -\gamma_{IJ} \omega^J\nonumber
\b
Integrating this equation along the homology cycles one obtains
\a
0=\p \oint _{a_i} dA_J &=& N_i^L~\gamma_{LJ}\label{null1}\\
0=\p \oint _{b_i} dA_J &=& M_i^L~\gamma_{LJ}\label{null2}
\b
Thus there are $2g+1$ eigenvectors of the matrix $\gamma$
with 0 eigenvalue,
taking into account that $\gamma_{J{g\over 2}}=0$
as a consequence of
$A_{g\over 2}\equiv 1$.

{}From (\ref{null1}) and (\ref{null2}) one has
\a
d(N_i^LA_L) &=& N_i^L~\gamma_{KL}~ \omega^K =0\nonumber\\
d(M_i^LA_L) &=& M_i^L~\gamma_{KL}~ \omega^K =0\nonumber
\b
Therefore $N_i^LA_L$ and $M_i^LA_L$ are constant (outside
$\P_\pm$).\par
The delta function property is introduced through
\a
\p \oint_{C_\tau}\Delta({\scriptstyle Q},{\scriptstyle Q}')
\psi({\scriptstyle Q})=\psi({\scriptstyle Q}')\nonumber
\b
for any one-form $\psi$, where
\a
\Delta({\scriptstyle Q},{\scriptstyle Q}') = A_J({\scriptstyle Q})
\omega^J({\scriptstyle Q}') \nonumber
\b
Finally let me introduce for later convenience the constants
${\alpha^L}_{JK} $, defined through
\a
{\alpha^L}_{JK} = \p \oint A_JA_K\omega^L
\label{alpha}
\b
In genus $0$ we have
\a
{\alpha^L}_{JK}= \delta_{L,J+K}, \quad\quad \gamma_{JK}=
{\scriptstyle K}
\delta_{J+k,0}\nonumber
\b

{}~\\
\section{The KP hierarchy.}

\indent

Let us recall the basic features of the KP hierarchy. A convenient way
of introducing it makes use of the pseudodifferential operator (PDO)
formalism (see \cite{dickey}): without even referring to any
hamiltonian
structure, the KP flows can be defined via the
pseudodifferential Lax
operator $L$:
\a
L& =&\partial + \sum_{n=0}^\infty u_n\partial^{-n-1}
\label{laxop}\b
where the $u_i$ are an infinite set of fields depending on the
spatial coordinate $x$ and the time parameters $t_k$.
The
different flows are defined through the position \begin{eqnarray}
{\partial L\over \partial t_k}& = &[ L, {L^k}_+]
\label{laxflow}
\end{eqnarray}
where $k$ is a positive integer and ${L^k}_+$ denotes the purely
differential part of the operator.\\
The quantities
\a
F_k &=& <L^k>
\label{fint}
\b
are first integral of motions for the
flows
(\ref{laxflow}) (the symbol $<A>$ denotes the integral of the residue,
$<A> = \int dw a_{-1} (w)$ for the generic pseudodifferential
operator\\
$A= ... + a_{-1} \partial^{-1}+... $ .
\par
The basic requirement of the
pseudodifferential calculus lays on the commutation rule
\a
\partial^{-1} f&=&f \partial^{-1} +\sum_{r=1}^{\infty}
(-1)^r f^{(r)} \partial^{-r-1}
\b
(where $f^{(r)} \equiv \partial^r f $),
together with the standard properties of the derivative (Leibniz
rule, etc.).
In order to formulate the KP hierarchy in higher genus, $L$ should be
regarded as a covariant operator, mapping tensors into tensors, and
it must be checked that the pseudodifferential calculus with
covariant derivatives works as in the standard case.\par
Let us denote as
$f_{\lambda} ={\overline f}_{\lambda}(z)(dz)^{\lambda} $ a
$\lambda$-weight tensor
(from now on, overlined quantities will always denote the expression
in local coordinates).
The covariant derivative, mapping tensors of weight $\lambda$ into
tensors of weight $\lambda +1$ is introduced through the position
\a
{\cal D} &=& dz ( \partial_z -\lambda {\overline \Gamma} (z))
\b
where $\Gamma$ is a connection, transforming under
reparametrizations as the logarithmic derivative of a $1$-form.\\
The space of connections is an affine space, which means that
every
connection can be expressed as the sum of a given
logarithmic derivative of reference plus a holomorphic (outside
$P_\pm$) $1$-form. Even if any connection can be used to form
a covariant derivative, neverthless, the PDO formalism
seems inconsistent
unless $\Gamma $ can be expressed as a logarithmic derivative of a
$1$-form, no matter which\footnote{This has to do with the fact that
``inverting" a generic covariant derivative
leads to non-single valued integrals.}.
Therefore I will consider the PDO
formalism only for the restricted class of connections which
can be represented as
\a
{\overline\Gamma} &=& {\textstyle{\partial {\overline \omega}\over
{\overline\omega }}}
\label{omegacon}
\b
$\omega $ can be chosen arbitrarily; different choices will provide
different dynamics. There is however a uniquely defined canonical
choice for $\omega$: in order to allow
the identification of the dynamics with the euclidean evolution on
$\Sigma$ only the choice $\omega \equiv {\omega^{g\over 2}} $
(the abelian
differential of the third kind normalized in such a way that its
integrals along the homology cycles are purely imaginary) is
allowed. This point has been discussed in \cite{bm},
so that I will not
come back any longer on it.
\par
Notice that $\omega$ with respect to the connection defined out of it
is a covariantly constant $1$-form, namely
it satisfies
\a
{\cal D}_{\Gamma} \omega &=& 0
\b
Therefore $\omega$ can be used
to construct covariantly constant tensors at any order,
turning
any constant into a
covariant constant having the wished tensorial properties.\par
At this point we have to introduce the ``inverse" operator
with respect to the covariant derivative. To do that let
us introduce the
covariant integral $\int_{\Gamma}$ associated to the connection
$\Gamma$; the covariant integral maps tensors of weight $\lambda$
into tensors of weight $\lambda -1$. The right choice for
$\int_{\Gamma}$ is the following:
\a
\int_{\Gamma} &=& {\omega}^{\lambda} (Q) \cdot \int_{Q_0}^Q
{\omega}^{\lambda - 1}({\overline Q})\cdot
\b
Here $Q_0$ is a given reference point, the only requirement being
$Q_0 \neq P_{\pm}$. It is a straightforward check to prove
that $\int_{\Gamma}$ has the right properties:
one gets indeed
\a
{\cal D}_{\Gamma} \cdot \int_{\Gamma} &=& {\bf 1}
\b
and
\a
 \int_{\Gamma}{\cal D}_{\Gamma} f^{\lambda} (Q) &=&
f^{\lambda}(Q) - \omega^{\lambda} (Q){ f^{\lambda}(Q_0) \over
\omega^{\lambda} (Q_0)}
\b
where the last quantity on the r.h.s.
is a covariantly constant $\lambda$-tensor.
With the above definitions covariant tensors and covariant integrals
can be used as ordinary derivatives and integrals: in particular it is
possible to integrate by parts; moreover the covariant integral
of a totally covariant derivative along any closed
contour ${\cal C}$ is
always vanishing:
\a
{\oint}_{{\cal C},\Gamma}{\cal D}_{\Gamma} f &=& 0
\b
Therefore it makes sense to introduce the symbol ${\cal D}^{-1}$
and the standard rules of the PDO calculus can be applied as before.
Notice that with the above definitions one can take contour integrals
for tensors of any order.
\par
{}From now on, to simplify notation, covariant integrals will be denoted
with the same symbol as ordinary integrals: no confusion will arise,
the context will say which is which.
\par
{}
The covariant KP hierarchy is defined by replacing the
ordinary derivative with the covariant one in
(\ref{laxop}).
Taking into account that $L^k$ is a $k$-th differential, the covariant
flows of the KP hierarchy can be introduced through
\begin{eqnarray}
\omega^k{\partial L\over \partial t_k}& = &[ L, {L^k}_+]
\label{lax2}
\end{eqnarray}
(no summation in the l.h.s. of course).\\
The first integrals of motions
are obtained from (\ref{fint}) by replacing the integrals
with covariant ones and taking $C_{\tau}$ as contour integration.
Due to the presence of the poles in ${\textstyle{1\over \omega}}$,
different choices of $\tau$ lead to different values for the first
integrals. Once fixed a particular ${\overline \tau}$,
the curve $C_{\overline\tau}$ can be regarded on the same foot as
the line integrals in the flat case.\par
\par
Up to now no Poisson brackets structure has been introduced.
However once
the ``covariantization rules" are at disposal,
it is quite immediate to introduce the set of
Poisson brackets
by repeating the steps of \cite{dickey}. In order to leave this
paper at a
reasonable letter-size, all that will not be discussed explicitly,
rather in the next
section the Poisson brackets dynamics will be introduced through
the perhaps
more interesting framework of matricial KP, and the connection
between matricial
and scalar KP will be pointed out.
\par
Here let us simply recall that the standard Drinfeld-Sokolov reductions
of the KP hierarchy can be obtained by imposing the constraint,
consistent
with the flows (\ref{laxflow})
\a
L^n &=& {L^n}_+\equiv L_n
\b
(with $n$ positive integer $=2,3,...$), which tells that the $n$-th
power
of $L$ is a purely differential operator.
For $n=2$ one gets the higher genus KdV equation
($ L_2 = {\cal D}^2 + T$); for $n=3$ the higher genus
Boussinesq equation ($L_3 ={\cal D}^3 + U{\cal D} + V$). \par
The Poisson brackets structure can be introduced also through
free-fields Miura maps, namely by representing
the $n$-th differential operator
through the position
\a
L_n & =& ({\cal D} -\Phi_1)({\cal D} -\Phi_2)...({\cal D}-\Phi_n)
\b
supplemented by the condition $\Phi_1 +\Phi_2+...+\Phi_n =0$.
The $n-1$ independent fields $\Phi_i$'s are $1$-forms satisfying the
free-fields algebra on Riemann surfaces, i.e. the higher genus
Heisenberg algebra (\ref{At}):
\a
\Phi_i (Q) &=& \alpha_{J,i}\omega^J (Q)\nonumber\\
\{\Phi_i(Q),\Phi_j(Q') \} &=&
-\delta_{ij}\gamma_{IJ}\omega^I(Q)\omega^J(Q')
\b
for $i,j = 1,...,n-1$.\par
The KdV field $T(Q)$ in $L_2$ can be represented as
\a
T(Q) &=& {\cal D} \Phi (Q) - \Phi^2 (Q)
\b
It is a straightforward check to show
that when $\Phi(Q)$ is assumed to
satisfy the higher genus Heisenberg algebra
(\ref{At}), then the Poisson brackets algebra of the
higher genus momenta
\a
L_I &=& {{\textstyle 1\over 2\pi i}}\oint e_I T
\b
corresponds to the Krichever-Novikov algebra with central charge $c=1$.
The cocycle will be fixed by the choice of the connection
in ${\cal D}$,
however different connections will give rise to different cocycles
belonging to
the same cohomology class. The value of the central charge is
uneffective
at the classical level since it can always be
rescaled to any non-vanishing
value via field redefinitions and Poisson brackets rescalings. \par
Similarly the two fields $\Phi_{1,2}$ in
$({\cal D} -\Phi_1)({\cal D} -\Phi_2)({\cal D}+\Phi_1 +\Phi_2)     $
will give rise a higher genus $W_3$ algebra structure for the
Boussinesq fields $U$ and $V$.
The momenta for the $3$-tensor $V$ should be taken with respect to the
basis ${f^{-2}}_I$ of weight $-2$ tensors:
$V_I = {\textstyle{1\over 2\pi i}}
\oint {f^{-2}}_I V$.
When $T$ (and respectivly $U,V$) satisfy the
KdV (Boussinesq) equation on Riemann
surface, the fields $\Phi$ ($\Phi_{1,2}$) satisfy the corresponding
higher genus mKdV (modified Boussinesq).
The extension of this structure to generic values of $n$ is immediate.
\\
{\quad}\\

\section{The NLS Hierarchy from Matrix KP.}

\indent

In this section I will show how to put on Riemann surfaces the AKS
approach to the hierarchies based on matrix-type Lax operators
\cite{genkdv}
and to connect it with the scalar KP;
the scheme here adopted has been presented in \cite{top3}
and is part of a
more developed forthcoming paper on this topic.
\par
One starts with the matrix Lax operator
\a
{\cal L} &=& \partial_x + J(x) +\lambda K
\b
where $J(x)$ are currents valued in some finite Lie algebra $\cal G$,
$\lambda$ is a spectral parameter and $K$ is a constant
element in ${\cal G}$.
The Kac-Moody current algebra ${\hat {\cal G}}$ is one of the
Poisson brackets
structure for the above system (the one we are interested in).
The generalized
Drinfeld-Sokolov hierarchies are obtained (see\cite{genkdv})
by assuming $K$ being a
regular element for the loop algebra ${\tilde{\cal G}}$, with $\lambda$
as loop parameter. The regularity condition means that
\a
{\tilde{\cal G}} &=& Ker K \oplus Im K
\b
where the action is the adjoint one.
\par
Under the above assumptions it is possible to introduce a
grading $deg$
on the elements of the loop algebra, defined in such a way that
$deg(\lambda K)=1$.Different gradings induce different hierarchies:
the principal grading, which associates the grade one to
the simple positive
roots of the algebra, provides the standard DS hierarchies. Another
interesting kind of gradings is furnished by the homogeneous one
($ deg \equiv {\textstyle{d\over d\lambda}}$).
The NLS hierarchy is obtained
by taking the homogeneous grading w.r.t.
the ${\cal G} \equiv sl(2)$ algebra
having $H$ as Cartan generator and $E_\pm$ as roots. In this case
$K=H$ is a regular element.\par
The main property of the above operator can be stated as follows:
there exists
an adjoint transformation ${\cal L} \mapsto {\cal L}_\alpha =
adj(\alpha){\cal L}$,
\a
{\cal L}_{\alpha} &=& {\cal L}+ [\alpha,{\cal L} ] +
{\textstyle{1\over 2}}
[\alpha,[\alpha,{\cal L}]] +...
\b
which preserves both the Poisson brackets and
the monodromy invariants. Under
the condition of regularity for $K$ it is possible
to uniquely determine, with
an iterative procedure, the local fields
$\alpha (x) \in Im K$ ($\alpha (x)$
expanded in the components with negative gradings only), such that the
transformed operator ${\cal L}_\alpha$ is diagonal:
\a
{\cal L}_\alpha &=& \partial_x +R(x) +\lambda K
\b
where $R(x)\in Ker K$ is expanded over the non-positive
grading components
and can be iteratively computed.The diagonal character of
the transformed
operator makes possible to compute the monodromy invariants.
The different components
of $R(x)$ provide the tower of hamiltonian densities.\par
To be explicit, in the NLS example we have
\a
\alpha (x) &=& \lambda^{-1}\alpha_1 +\lambda^{-2}\alpha_2 +...
\b
with
\a
\alpha_j &=& \alpha_{j,+} E_+ + \alpha_{j,-}E_-
\b
and
\a
R(x) &=& R_0 H + \lambda^{-1} R_1 H +\lambda^{-2}R_2 H +...
\b
At the lowest order, with easy computations, we get
\a
R_0 &=& J_0\nonumber\\
R_1 &=& 2J_+J_-\nonumber\\
R_2 &=& J_+\Delta J_- - J_- \Delta J_+
\label{integr}
\b
(where $\Delta$ is the covariant derivative w.r.t. the Kac-Moody
${\hat {U(1)}}$
subalgebra:
${\Delta J_\pm} = (\partial \pm 2 J_0)J_\pm$, see \cite{toppan}).
$R_2$ provides the hamiltonian density for the
Non-Linear-Schr{\"{o}}dinger
equation with respect to the Kac-Moody Poisson brackets.
The hamiltonians of the infinite tower are all in
involution with respect to
the Poisson brackets structure. Moreover
it is easily shown
by induction that every hamiltonian density $R_i$ with $i>0$
has vanishing
Poisson brackets with the current $J_0$, i.e. it belongs to the
${\hat {U(1)}}$ coset.\par
Consistent reductions of the previously considered
Lax operator $L$ of the scalar KP
hierarchy
can be recovered from the matrix KP ${\cal L}$ operator with
the following procedure:
let us take for simplicity the $sl(2)$ case in the
fundamental representation,
then if we solve the matrix equation
\a
\left(\partial {\bf 1} + \left(\begin{array}{cc}
J_0 & J_+\\
J_-& -J_0{}
\end{array}
\right)\right) \left(\begin{array}{c}\Psi_+\\
\Psi_-
\end{array}\right) &=& 0
\b
for, let'say, the $\Psi_-$ component and allow
inverting the derivative operator, we can plug the result
into
the equation for the $\Psi_+$ component, obtaining
\a
(\Delta + J_- \Delta^{-1} J_+)\Psi_+ &=& 0
\b
The operator $L=\Delta + J_- \Delta^{-1} J_+$ provides a
consistent reduction
of KP (the one associated with the NLS equation, see also
\cite{toppan2}).
A Poisson brackets structure for $L$ is induced and coincides
with the
Kac-Moody Poisson brackets for the matrix KP operator.
Analogous steps can be performed in the general case
as well (choice of different
affine Lie algebras and different representations).\par
In order to define the matrix KP hierarchy on Riemann surfaces,
we have to
promote the Lax operator ${\cal L}$ to be a covariant operator mapping
$\lambda$
tensors into $\lambda + 1$ tensors. Therefore the following
substitutions
should
be made: the ordinary derivative should be replaced, as before, by the
covariant
derivative ${\cal D}$; the constant regular element $\lambda K$
should now
be regarded as a covariantly constant $1$-form, therefore
$\lambda K \mapsto \omega\lambda K$, with $\omega $ introduced in
(\ref{omegacon}).
As for the term containing the currents $J_i$'s, it should be given by
$1$-forms
taking values in the Lie algebra ${\cal G}$; their Poisson brackets
should
be consistent with the requirement that the $J_i$'s live in the
Riemann surface
$\Sigma$, and as a consequence it must be provided by
the higher genus Kac-Moody algebra
(which plays for Kac-Moody the same role as the
Krichever-Novikov algebra
for Virasoro); a presentation of this algebra is given in
${\cite{bt}}$.\par
To be definite, let us treat here explicitly the derivation of the NLS
hierarchy, the generic

case will follow immediately with trivial modifications.\\
We have now ${\cal G} {\equiv} \{H, E_\pm\}$, and commutation relations
\a
{} [ H, E_\pm ] &=& \pm E_\pm \nonumber\\
{}[E_+, E_-] &=& 2 H
\b
${\cal L}$ is given by the covariant operator
\a
{\cal L} &=& {\cal D} +\lambda H \omega + E_\pm J^{\pm} +HJ^0
\b
The fields $J^{\pm}(Q), J^0(Q)$ are $1$-forms which can be expanded
in their
KN-modes as $ J^i = {J^i}_I\omega^I$. In a convenient normalization
their Poisson brackets algebra can be expressed as
\a
\{J^0(Q), J^0(Q')\} &=& d_{Q'}\Delta(Q',Q) \nonumber\\
\{ J^0(Q), J^\pm (Q')\} &=& \pm 2 \Delta (Q',Q) J^\pm (Q')\nonumber\\
\{ J^+(Q),J^-(Q')\} &=&
\Delta (Q',Q)J^0(Q') +{\textstyle {1\over 2}}d_{Q'}
\Delta(Q',Q)
\label{sl2hg}
\b
and, in terms of the modes:
\a
\{{J^0}_I,{J^0}_J\}&=&-\gamma_{IJ}\nonumber\\
\{{J^0},{J^\pm}_J\} &=& \pm 2 {\alpha^K}_{IJ}{J^\pm}_K\nonumber\\
\{{J^+}_I,{J^-}_J\} &=& {\alpha^K}_{IJ} {J^0}_K -{\textstyle{1\over 2}}
\gamma_{IJ}
\label{modes}
\b
where the constants in the r.h.s. are defined in section $2$.
The diagonalization procedure outlined above in the flat case can
be repeated
to get the covariant diagonal operator ${\cal L}_{\alpha}$.
The hamiltonians are the covariant versions of those of (\ref{integr}),
the integration contour being $C_{\tau}$. They are in involution
with respect to the higher genus Poisson brackets.
At the lowest order we have as hamiltonian densities
\a
R_0 &=& J^0\nonumber\\
R_1 &=& 2J^+J^-\nonumber\\
R_2 &=& J^+{\hat\Delta} J^- - J^- {\hat \Delta} J^+
\label{nlshg}
\b
In (\ref{nlshg},c) the derivative ${\hat\Delta}$, covariant
with respect to both reparametrizations and the Heisenberg
subalgebra (\ref{modes},a) appears.
In general, if ${V^\lambda}_q$ is a $q$-charged $\lambda$-tensor, which
means that
\a
\{ J^0(Q), {V^\lambda}_q(Q')\}&=& q \Delta (Q,Q'){V^\lambda}_q (Q)
\b
the covariant derivative ${\hat\Delta}$ acts as follows
\a
{\hat\Delta}{V^\lambda}_q &=& ({\cal D} - qJ^0){V^\lambda}_q
\equiv {V^{\lambda +1}}_q
\b
and maps ${V^\lambda}_q$ into a $\lambda +1$ tensor of
definite charge $q$.\\
Here
\a
{\hat \Delta} J^\pm&=& ({\cal D}\mp 2 J^0) J^\pm
\b
As before, all the hamiltonian densities apart from $R_0$ belong to
the coset with respect to the
higher genus Heisenberg algebra (\ref{modes},a).\par
The two-components NLS equation is derived from the third hamiltonian
$\oint R_2$ and reads as:
\a
\omega^2 {\partial\over\partial t} J^\pm &=&
\pm{\hat\Delta}^2J_\pm \pm 2 (J^+J^-)J^\pm
\b
(the equation relative to $J^0$ is immediately solved to give
$J^0 \propto \omega$).
\par
As in the genus zero case we can derive the scalar form of the
NLS hierarchy, which coincides with a particular (non-canonical)
reduction of KP. In its more compact form it can be
represented as
\a
L &=&{\hat \Delta} + J^-{\hat\Delta}^{-1} J^+
\label{laxred}
\b
where the pseudodifferential calculus is defined for the
totally covariant operator ${\hat \Delta}$. The justification of this
procedure is contained in \cite{toppan2}. Obviously we can rexpress
(\ref{laxred}) as a pseudodifferential operator in terms of the
derivative ${\cal D}$; in this case we mantain the manifest
covariance w.r.t. the
diffeomorphisms, but not w.r.t. the Heisenberg subalgebra: as a result
the computations are more involved and less transparent.\par
The appearance of a ${\cal W}$ algebra is due to the fact that every
hamiltonian density is a polynomial function
of the composite fields
\a
W_n &=& {\hat\Delta}^nJ^+\cdot J^-
\b
($n$ non-negative integers) and their covariant (${\cal D}$)-derivatives.
The same considerations as in \cite{toppan} hold: the higher order
fields ($n\geq 2)$ are algebraic functions of
$W_{0,1}$ and their ${\cal D}$-derivatives: the algebra generated by
$W_0$,$W_1$ is a higher genus ${\cal W}$ algebra
which closes in a rational way, i.e. in the r.h.s.
terms proportional to the higher order fields appear.
  It is convenient to express
such algebra in terms of the fields \par
$ T= {\textstyle{1\over 2}} J^+J^- $ and $\Psi = {\hat D} J^+ \cdot J^-
-{\hat \Delta}J^- \cdot J^+ $. \par
$T$ plays the role of a stress-enery tensor
having (only classically) a vanishing central charge, while $\Psi$ is
a primary field w.r.t. $T$ having conformal dimension $3$.
The algebra is explicitly given by
\a
\{T(Q),T(Q')\} &=& 2 \Delta^1  \omega (1)
T -\Delta \omega (2)
T^{(1)}\nonumber\\
 {}\{ T(Q), \Psi (Q')\} &=& 3 \Delta^1
\omega (1)
\Psi - \Delta \omega (2)
\Psi^{(1)}\nonumber\\
{}\{\Psi (Q),\Psi (Q')\} &=& 2\Delta^3 T -3 \Delta^2 \omega (1) T^{(1)}
+ 2 \Delta^1 \omega (2) A -\Delta \omega (3) (A^{(1)}-
{\textstyle {1\over 2}} T^{(2)})
\b
where the compact notation
\a
{{\cal D}_Q}^n\Delta (Q,Q') &\equiv& \Delta^n\nonumber\\
{\cal D}^n (T, \Psi) &\equiv& T^{(n)} , \Psi^{(n)}\nonumber\\
{\textstyle {\omega^n (Q)\over\omega^n (Q')} }&\equiv & \omega (n)
\nonumber\\
A &=& 8 W_2 -4\Psi^{(1)} +4 T\cdot T -2 T^{(2)}
\b
has been used. In the r.h.s. all the fields are evaluated in $Q'$.
One should notice on the r.h.s. the appearance of the fields
$W_2$, which is an algebraic functions of $T$, $\Psi$.
Since $T$ and $\Psi$ are chargeless, it does not matter if the r.h.s.
is expressed in terms of the ${\hat\Delta}$ or the ${\cal D}$
covariant derivatives.
It should be pointed out that the existence of the covariantly constant
$1$-form $\omega$ allows to have the right tensorial
properties for the above algebra in both the points $Q$, $Q'$.\par
To end this section, let us introduce the free-fields classical
Wakimoto representation in higher genus: it provides the NLS
extension of the Miura map and allows defining
the modified NLS hierarchy (see \cite{toppan2}).
It is realized in terms of the $1$-form $p$ and by the
coupled (respectivly ($1\backslash 0$) tensors) $\beta\backslash \gamma$.
A convenient way of representing such fields is the following:
\a
p(Q) &=& \alpha_I\omega^I\nonumber\\
\beta(Q) &=& \beta_I\omega^I\nonumber\\
\gamma (Q) &=& \gamma (Q_0) +\int_{Q_0}^Q \gamma_I\omega^I
\b
The assumed Poisson brackets in higher genus are the free ones, namely
\a
\{p(Q), p(Q')\} &=& d_{Q'}\Delta(Q',Q) \nonumber\\
\{ \beta(Q), \gamma (Q')\} &=& \Delta (Q',Q)
\b
and vanishing otherwise.\\
In terms of the modes they read
\a
\{{\alpha}_I,{\alpha}_J\}&=&-\gamma_{IJ}\nonumber\\
\{{\beta}_I,\gamma (Q_0)      \} &=& A_I(Q_0)\nonumber\\
\{{\beta}_I,{\gamma}_J\} &=& -
\gamma_{IJ}
\b
Since $\gamma$ must be univalued on $\Sigma$, the requirements
\a
0={\hat\gamma}_i &=& \gamma_I {N^I}_i\nonumber\\
0={\check \gamma}_i &=& \gamma_I{M^I}_i
\label{requ}
\b
should be imposed ($ {N^I}_i$, ${M^I}_i$ are given in
(\ref{Ni},\ref{Mi})). \par
It is a remarkable feature
of the Heisenberg algebra on higher genus that ${\hat\gamma}_i $,
${\check\gamma}_i $ have vanishing
Poisson brackets with respect to any element of the algebra,
and therefore
(\ref{requ}) can be imposed without further constraints.
The algebra (\ref{sl2hg}) can be reproduced through the

identification
\a
J^+ (Q) &=& \beta (Q)\nonumber\\
J^0 (Q) &=& (p -2\beta\gamma)(Q)\nonumber\\
J^- (Q) &=& ( p\gamma -\beta\gamma^2 +
{\textstyle {1\over 2}} d\gamma ) (Q)
\label{waki}
\b
as it can be straightforwardly checked.
The modified NLS equation can be derived, by inserting (\ref{waki})
in $R_2$: we obtain the coupled system for $\beta , \gamma$
\a
\omega^2 {\dot \beta} &=& {\cal D}^2 \beta + 2\beta^2{\cal D}\gamma -
2\beta^3\gamma^2\nonumber\\
\omega^2{\dot\gamma}&=& {\cal D}^2\gamma -2\gamma^2{\cal D}\beta -
2\gamma^3\beta^2
\b
which generalizes the genus zero results. Here the constraint
$J^0 =0$, which
is consistent with the equations of motion, is set to get rid
of the field
$p$ in the above equations.
{}\\
{}\\
\noindent

{\Large {\bf Conclusions}}

\indent

The KP hierarchy and its reductions have recently being
investigated in connection with the matrix model formulation of the
$2$-dimensional gravity. Starting from the standard KP hierarchy
there are several possible ways of generalizing it: by enlarging its
algebraic setting allowing supersymmetry; by quantizing it (it
could be speculated that such quantum versions are linked to some
string-field theory); by $q$-deforming it, which leads to the
connection with quantum group. In this paper I have shown
an appropriate framework to put it on Riemann surfaces, which became
popular in physicists' literature due to the Polyakov
formulation of the string perturbation theory.
It is likely that the formulation here proposed be relevant
for investigating the genus expansion of matrix models. This will be
left for future investigations.
{}~\\~\\

\noindent
{\large{\bf Acknowledgements}}
{}~\\~\\
I wish to express my thanks to Prof. M. Tonin and to the Physics
Departement of the Padua University for their kind hospitality.
{}~\\
{}~\\
{}\vfill\eject

\end{document}